\documentstyle[12pt]{article}
\begin{document}
\title{The string model of the  Cooper pair
in the anisotropic superconductor}
\date{}
\author{V.D.Dzhunushaliev
\thanks{E-mail address: dzhun@freenet.bishkek.su}}
\maketitle
\begin{center}
Theoretical physics department, the Kyrgyz State National
University, 720024, Bishkek, Kyrgyzstan
\centerline{PACS numbers:74.90.+n; 11.27.+d}
\end{center}
\begin{abstract}
\par
The analogy between the Cooper pair in high temperature
superconductor and the quark-antiquark pair in quantum
chromodynamics (QCD) is proposed. In QCD the nonlinear chromodynamical
field between a quark and an antiquark is confined to a tube.
So we assume that there is the strong interaction between phonons
which can confine them to some tube too. This tube is described
using the nonlinear Schr\"odinger equation. We show that it has an
infinite spectrum of axially symmetric (string) solutions with
negative finite linear energy density. The one-dimensional nonlinear
Schr\"odinger  equation  has a finite spectrum (hence, it has
a steady-state) which describes the Cooper pair
squezeed between anisotropy planes in the superconductor. It is shown
that in this model the transition temperature is approximately 45 K.
\end{abstract}
\par
The quantum theory of solids has much in common with
quantum field theory. For example, it is well-known that there
exists the analogy between Abrikosov's magnetic curl in the theory of
superconductivity  and the tube of the gluon field between
quarks. The quanta of the corresponding interactions (i.e. photons,
intermediate vector bosons, gluons on the one hand and phonons on the
other hand) have common properties too. A detailed comparison
of the above-mentioned theories is made in \cite{dzh1}. Nevertheless
there are some essential distinctions between these theories:
the first is a gauge theory and the second is a lattice theory.
In this article we try to apply the similarity of these theories.
\par
Both lattice calculations in quantum chromodynamics (QCD)
and experiments in nuclear physics indicate that the chromodynamical
field between quarks due to the strong interaction between gluons
is confined to a thin tube stretched between quarks.
Perhaps, the appearance of the tube explains the quark confinement
phenomena, that is the impossibility of their existence in the free
state. In QCD such tubes filled with chromodynamical field (see, for example,
\cite{nus}, \cite{ols}) have been under investigation for a long time.
It may turn out that there is some kind of strong
interaction between phonons in superconductors.
Such a possibility is discussed in \cite{kri} where  the  inclusion  of
vertex corrections to the calculation of $T_c$ is studied. Then by analogy
with QCD one can suppose that in this case the phonons are also
confined to a thin tube stretched between the Cooper pair of
electrons. In this case the Cooper pair energy will consist of
two terms: the energy of the Cooper pair calculated by
Cooper, and the energy of the phonon tube (PT) which we will try to
calculate in this paper. The presence of the second term (PT energy)
automatically leads to a rise in the superconducting transition
temperature.
\par
In \cite{orl}, the Rokhsar - Kivelson model on  the  square  lattice
has been studied. This system is a quantum dimer model on a square
lattice,   proposed   as   an   effective   theory   of    layered
superconductors. It has been shown that  the  Rokhsar  -  Kivelson
model is equivalent to a gas of transversally  oscillating  "hard"
strings.
\par
The colour tube in QCD has a nonzero chromodynamic field. Analogously
it can be supposed that the PT has a nonzero "phonon" field, that is a
lattice deformation between Cooper pairs. In  \cite{net}  it  is
proposed  that
"$\ldots$ the Cooper pair is bound, not only by phonons
in  the  usual  way but by static lattice deformations as well$\ldots$".
\par
The PT between the Cooper electrons is a quantum object,
as is also the case with the gluon tube between quarks. We will try to
describe this object using a nonlinear Schr\"odinger (NS) equation.
In favour of this statement we will give the following arguments
\cite{dzh1}. Let us take a small piece of the PT. According
to Feynman the transition amplitude of this piece from the initial point
to the final point is the sum over all paths (i.e. the path integral).
This means that the different PT pieces can pass through any space
point simultaneously. We suppose that this leads to a nonlinear
quantum self-action of the
PT wave-function which is analogous to the nonlinear term in the
Ginzburg-Landau (GL) equation describing the wave-function of the
Cooper pairs. We write down the transition amplitude $<b~|~a>$
from the point $\vec r_a$ to the point $\vec r_b$:
\begin{equation}
<b\mid a> = \int D\vec r[t]
\exp \left\{\frac{i}{\hbar}\int\limits _{a}^{b}
\left(E_{k}-U\right) dt \right\},
\label{1}
\end{equation}
where $E_{k}$ is the kinetic energy and $U$ is the potential energy.
According to the above mentioned arguments we suggest that
\begin{equation}
U = V - \lambda \mid\Psi\mid ^2,
\label{2}
\end{equation}
where $V$ is the external potential; $\lambda$ is some constant
defining the quantum self-action of the PT wave function
$\Psi (\vec r,t)$. Actually  we supposed here that all the points
of the PT are described by the one and the same
wave function $\Psi(\vec r,t)$. Following the ordinary
method we can obtain the NS equation for the wave function of the
chosen PT piece:
\begin{equation}
i\hbar \frac{\partial\Psi}{\partial t} =
\left[-\frac{\hbar ^2}{2\mu}\nabla ^2 +
V - \lambda \mid\Psi\mid ^2\right]\Psi,
\label{3}
\end{equation}
where $\mu$ is the mass of the considered PT piece.
\par
Now we can proceed to the stationary state equation using the
following substitution:
\begin{equation}
\Psi \left( \vec r,t \right)=
\exp\left(\frac{iat}{\hbar}\right)\psi \left(\vec r \right),
\label{4}
\end{equation}
where $a$ is some constant. As a result we obtain the following
stationary NS equation:
\begin{equation}
-a\psi = \left(-\frac{\hbar ^2}{2\mu}\nabla ^2
- \lambda \mid\psi\mid ^2 \right)\psi,
\label{4a}
\end{equation}
where we equated the exterior potential to zero $(V=0)$. Let us compare
this equation with the stationary GL equation:
\begin{equation}
\frac{1}{\eta}\psi = \left(-\frac{\hbar ^2}{2m}\nabla ^2 +
\frac{1}{\eta n_e}\mid\psi\mid ^2\right)\psi,
\label{4b}
\end{equation}
where $m$ is the mass of the Cooper electrons pair;
$\eta=7\zeta (3)E_F/8\pi ^2 E^2_c$; $E_F$ is the Fermi energy;
$E_c$ is the transition temperature in energy units;
$\zeta$ is the Riemann zeta function; $\psi$ is the
order parameter; $n_e$ is the concentration of electrons
in the normal phase. It is necessary to note that in \cite{vons}
the authors proved the correctness of the GL equation as applied to
high-$T_c$ superconductivity. Comparing these two expressions,
we see that they differ only in the sign of the parameters
$(a, \lambda)$ and $\eta$. According to Eq.(\ref{4}) $a$ is some
energy, and the sign of
$\lambda$ testifies either to quantum attraction when
$\lambda >0$ or quantum repulsion when $\lambda < 0$. According to
Eq.(\ref{4a}) and Eq.(\ref{4b}) we will consider the parameter $\mu$
in (\ref{4a}) to be some mass. Therewith, the parameters
$a$ and $\lambda$ are still unknown. As the equations
(\ref{4a}) and (\ref{4b}) coincide  (up to a sign of parameters
$(a, \lambda)$  and $\eta$) we can say that the equation
(\ref{4a}) is a nonlinear Schr\"odinger equation of GL type
describing the PT.
\par
Henceforth we shall call the PT  a string and ignore
the transverse dimension of the PT. Nowadays strings
are the subject of fundamental investigations \cite{wit}. In
superstring theory the string is a basic object, like an elementary
particle. Thus, the NS equation (\ref{3}) is
a quantum equation for a conjectural nonlocal object,
i.e. the string (PT). Then we should make sure that the
equation really has a solution of string type.
\par
It should be noted that  there  are  various  attempts
to  formulate  effective
actions for ordinary Bardeen - Cooper - Schrieffer superconductors
based on nonlinear equations of GL type at T=0 (see, for  example,
\cite{ait}).
\par
We are looking for an axially symmetric solution for the infinite
string placed along the axis $z=0$ which has finite linear
energy density. It should be noted that in \cite{mar} the axially
symmetric GL equations are  solved
for a flux vortex carrying a longitudinal  current.  It  is  shown
that the GL equations are equivalent  to  the  radial  pressure  -
balance equilibrium relation in ideal magnetohydrodynamics. It  is
possible that this is similar to the chromodynamic case in  which  the
gluon vacuum compresses the colour force lines into a thin tube  or
even into a string between the quark and antiquark.
Thus the stationary NS equation takes the
following form:
\begin{equation}
y'' + \frac{y'}{x} = y\left(1-y^2\right),
\label{5}
\end{equation}
where we use the dimensionless variable
$x=(2a\mu /\hbar ^2)^{1/2}\rho$; the function
$y~=~(\lambda/a)^{1/2}\psi$ and $z$, $\rho$ and $\varphi$ are
cylindrical coordinates. This equation was investigated in
\cite{dzh1}. The results of  this investigation showed that
the equation has a discrete string spectrum of solutions which
are regular in the whole space. This means that solutions with linear
finite energy density exist only with some special boundary
values $y^*_n~=~y(x=0)$ ($n$ is the knot number of $y_n(x)$).
Every solution $y_n(x)$ is a separatrix in the phase space
$(y,y')$ and its asymptotic behaviour is the following:
\begin{equation}
y(x) \approx \frac{\exp (-x)}{\sqrt x}.
\label{5a}
\end{equation}
\par
It is obvious that in this case the solution drops so quickly to zero
that it leads to the finite linear energy density of the string.
The solutions are wound around stable focuses
$y_{1,2}=\pm 1$ with the remaining boundary values $y(0)\neq 0$. Therewith
the asymptotical behaviour $y(x)$ is:
\begin{equation}
y(x) \approx \pm 1 + C\frac{\sin \left(x\sqrt{2} + \phi _0\right)}
{\sqrt{x}},
\end{equation}
where $C$ and $\phi _0$ are constants.
\par
For the case with the asymptotic behaviour (\ref{5a}) we obtain
by numerical calculations the following values:
$y^*_0~=~2.206200\ldots$,
$y^*_1~=~3.331987\ldots$,
$y^*_2~=~4.150092\ldots$
and etc. It seems that $n~=~0,1,2\ldots\infty$.
The Eq. (\ref{3}) has the conservative integral (energy):
\begin{equation}
e_n = \int \left[\frac{\hbar ^2}{2\mu}
\mid\nabla\psi _n\mid ^2 - \frac{\lambda}{2} \mid\psi _n\mid ^4
\right]dV.
\label{6}
\end{equation}
\par
In our case it takes the following form:
\begin{equation}
e_n = \frac{\pi \hbar ^2 a}{\mu\lambda}
\int\limits_0^{\infty}\left({y'}_n^2 - \frac{1}{2}y_n^4\right)xdx.
\label{7}
\end{equation}
The numerical calculations show that the integrals:
\begin{equation}
I_n = \int\limits_0^{\infty}\left({y'}_n^2 - \frac{1}{2}y_n^4\right)xdx.
\label{8}
\end{equation}
have the following numerical values:
$I_1~=~-3.567\ldots$, $I_2~=~-8.451\ldots$,
$I_3~=~-13.278\ldots$ and etc. It is very likely that all $I_n<0$.
This means that such PT states are unstable due to the absence of any
minimum energy value. That is the situation in the 3-dimensional case.
\par
Let us consider the case when the Cooper pairs are
squeezed between two planes (as possibly occurs in the anisotropic
high-$T_c$ superconductors). In this case the general NS equation
takes the following 1-dimensional form:
\begin{equation}
y'' = y\left(1-y^2\right).
\label{9}
\end{equation}
Here a Cartesian coordinate system is introduced with
axis $z'$ perpendicular to the anisotropy plane;
the axis $y'$ is directed along the string and the axis
$x'$ is perpendicular to the string axis; by analogy with
(\ref{5}) the function $y=(\lambda /a)^{1/2}\psi$
and the dimensionless coordinate $x=(2a\mu /\hbar ^2)^{1/2}x'$
are introduced; $(')$ in (\ref{9}) means the derivative with respect to
$x'$. The stationary solution  of this equation is
a soliton:
\begin{equation}
y = \frac{\sqrt{2}}{\cosh x}.
\label{10}
\end{equation}
\par
Thus, the soliton describes here the PT stretched between
Cooper electrons and squeezed between anisotropy planes.
It should be noted that the soliton has been used in \cite{dav}
for an explanation of the high-$T_c$ superconductivity,
but there its physical sense is different. Also we can see
that the given solution is a separatrix and it has finite
energy density, whereas the solutions with $y(0)\ne \sqrt 2$
have an infinite linear energy density.
\par
Now we will calculate the linear energy density of the stationary
string (\ref{10}) (i.e. soliton) in the given anisotropic case:
\begin{equation}
e = \sqrt \frac{\hbar ^2 a^3}{2\mu\lambda ^2}
\int\limits_0^{\infty}\left(y'^2 - \frac{1}{2}y^4\right)dx.
\label{11}
\end{equation}
Replacing $y$ in (\ref{11}) by (\ref{10}) we obtain the following
value:
\begin{equation}
e = -\frac{4}{3}\sqrt \frac{\hbar ^2 a^3}{2\mu\lambda ^2},
\label{12}
\end{equation}
\par
In order to estimate the absolute value of the PT energy
we have to multiply (\ref{12}) by the tube length $l_0$
and the tube thickness $\delta x$. Assuming that
$\delta x$ coincides with the distance $\delta z$ between
the anisotropy planes of the superconductor we have:
\begin{equation}
|E| \approx \frac{2}{3}\frac{\hbar ^2 a}{\mu\lambda}l_0.
\label{13}
\end{equation}
Here we took into account that according to the definition
of the dimensionless coordinate $x$ the soliton thickness is equal to
$\delta x \approx \delta z \approx [\hbar ^2/(2a\mu )]^{1/2}$.
\par
We are not able to define the parameters $a$ and  $\lambda$ in
the present string model of the PT, but we can try to
take a guess about their values by comparing Eq.(\ref{4a}) with the
standard GL Eq.(\ref{4b}) for the Cooper pair. Thereafter
we may suppose that:
\begin{equation}
\frac{a}{\lambda}=n,
\label{15}
\end{equation}
where $n$ is some concentration. We can suppose that this is the
phonon concentration, $n=n_{ph}$. Then the value $E$ may be estimated
in the following way:
\begin{equation}
|E| \approx \frac{\hbar ^2}{\mu}n_{ph} l_0.
\label{17}
\end{equation}
Let us assume that $\mu$ is the total mass of the PT, and that it
can be expressed in term of the  mass of one phonon $m_{ph}$
and the PT volume
$V\approx l_0 \delta z^2$,
$\mu \approx m_{ph} n_{ph} l_0 \delta z ^2$. Thus:
\begin{equation}
|E| \approx \frac{\hbar ^2}{m_{ph}\delta z^2},
\end{equation}
we recall that $\delta z$ here is the distance
between the anisotropy planes.
\par
We estimate the phonon mass with the help of the kinetic
energy relation for the Debye phonon
$k_BT_D=m_{ph}v^2/2$, where $T_D$ is the Debye
temperature and $v$ is the phonon velocity i.e. the sound velocity.
According to Ref.\cite{al} we can put $T_D \approx 4\times 10^2K$,
$v~\approx ~4\times 10^3 m/s$. Then $m_{ph}\approx 0.7\times 10^{-27} kg$.
We take the following value $\delta z \approx 10\AA$,
and obtain the numerical value of the PT energy
$E \approx 3.1\times 10^{-21}J$, which increases the
superconducting transition temperature up
$\Delta T = E/k_B \approx 45 K$. Of course, this is an
approximate calculation. The accurate result can be
quite different from the one indicated here.
\par
In  this  connection  we  have  to  note  that  (\ref{4a})   is
applicable when T=0. Near the critical temperature the type  of  this
equation can be changed. (\ref{4a}) must  be  derived  from  microscopic
theory as is done when deriving phenomenological Ginzburg -  Landau
theory from microscopic Bardeen - Cooper - Schrieffer  theory.  In
this case it is possible to derive the temperature dependence
of  the  coefficients in (\ref{4a}).
\par
Finally we can comment that
in the suggested model of the Cooper pair in the anisotropic
superconductor the increase of the transition
temperature is due to the fact that the phonons are concentrated
into a tube
stretched between the Cooper electrons (as a consequence  of
the strong interaction between phonons). It is similar
in appearance to the chromodynamical tube between
quarks  (as is well-known, this tube leads to confinment
of quarks in QCD). Thus, we may say that the phenomena of
high-$T_c$ superconductivity and the quark confinement
are analogous to each other in the given model: they are based
on the appearance of the tube filled with one or another field
which is the carrier of the interaction. We emphasize that
the existence of strong interactions between the force quanta
is a necessary condition for appearance of such a tube. The
appearance of a PT explains why the measurements
of the energy gap using the tunneling method and
infrared spectroscopy give different results. The point
is that in the first case the complete destruction
of the Cooper pair takes places but in the second case the
absorbed photon activates the Cooper electrons to a state with
zero energy without destroying the PT. This
model also describes why the high-$T_c$ superconductors are
anisotropic. The reason is that the energy spectrum of
the PT is not limited from below in the nonanisotropic case.
\par
In summary we should note that the suggested model is
phenomenological and, hence, it cannot describe the reason
for the appearance of a PT. As a result of the above-mentioned
analogy we can assume that the PT formation is analogous to the
chromodynamical tube and is due to the essential nonlinearity
of the potential term in the corresponding Lagrangian.
\par


\begin{thebibliography}{20}
\bibitem{dzh1}
V.D.Dzhunushaliev, Superconductivity: physics, chemistry,
technique, v.7, N5, 767(1994).
\bibitem{nus}
S.Nussinov, Phys.Rev.D50, 3167(1994).
\bibitem{ols}
Dan LaCourse, M.G.Olsson, Phys. Rev.D39, 2751(1989).
\bibitem{kri}
H.R.Krishnamurty, D.M.Newns, P.C.Pattnail, C.C.Tsuei, C.C.  Chi,
Phys. Rev., B49, 3520(1994).
\bibitem{orl}
P.Orland, Phys. Rev., B49, 3423(1994).
\bibitem{net}
S.J.Nettel and  R.K.MacCrone,  Phys.  Rev.,  B47,  11360(1993);
S.J.Nettel and R.K.MacCrone, Phys. Rev., B49, 6395(1994).
\bibitem{vons}
S.V.Vonsovskii, M.S.Svirskii, Superconductivity: physics,
chemistry, technique, v.6, N9-10, 1787(1993).
\bibitem{wit}
M.E.Green, J.H.Schwarz, E.Witten, Superstring theory, v.1,2,
CUP, Cambridge,(1987).
\bibitem{ait}
I.J.Aitchison, P.Ao, D.J.Thouless,  X.M.Zhu,  Phys.  Rev.,  B51,
6531(1995); M.Stone, Int. J. Mod. Phys. B9, 1359(1995).
\bibitem{mar}
G.E.Marsh, Phys. Rev., B49, 450(1994).
\bibitem{dav}
A.S.Davydov, Hightemperature superconductivity, Kiev,
Naukova dumka, p.176(1990).
\bibitem{al}
P.B.Allen, Z.Fisk, A.Megliori, in book:
Physical properties of High-$T_c$ temperature superconductors,
editor D.M.Ginsberg, World Scientific, p.450(1989).
\end{thebibliography}
\end{document}